\documentclass[aip,reprint,amsmath,amssymb]{revtex4-1}
 
\usepackage{graphicx} 
\usepackage{dcolumn}
\usepackage{bm}

\begin{document}

\title{High repetition pump-and-probe photoemission spectroscopy based on a compact fiber laser system}

\author{Y.~Ishida}
\email[]{ishiday@issp.u-tokyo.ac.jp}
\affiliation{ISSP, University of Tokyo, Kashiwa-no-ha, Kashiwa, Chiba 277-8581, Japan}

\author{T.~Otsu}
\affiliation{ISSP, University of Tokyo, Kashiwa-no-ha, Kashiwa, Chiba 277-8581, Japan}

\author{A.~Ozawa}
\altaffiliation[Present address: ]{Max Planck Institute of Quantum Optics, Hans-Kopfermann-Str. 1, 85748 Garching, Germany}
\affiliation{ISSP, University of Tokyo, Kashiwa-no-ha, Kashiwa, Chiba 277-8581, Japan}

\author{K.~Yaji}
\affiliation{ISSP, University of Tokyo, Kashiwa-no-ha, Kashiwa, Chiba 277-8581, Japan}

\author{S.~Tani}
\affiliation{ISSP, University of Tokyo, Kashiwa-no-ha, Kashiwa, Chiba 277-8581, Japan}

\author{S.~Shin}
\affiliation{ISSP, University of Tokyo, Kashiwa-no-ha, Kashiwa, Chiba 277-8581, Japan}
\affiliation{CREST, Japan Science and Technology Agency, Tokyo 102-0075, Japan}

\author{Y.~Kobayashi}
\affiliation{ISSP, University of Tokyo, Kashiwa-no-ha, Kashiwa, Chiba 277-8581, Japan}
\affiliation{CREST, Japan Science and Technology Agency, Tokyo 102-0075, Japan}


\begin{abstract}
The paper describes a time-resolved photoemission (TRPES) apparatus equipped with a Yb-doped fiber laser system delivering 1.2-eV pump and 5.9-eV probe pulses at the repetition rate of 95~MHz. Time and energy resolutions are 11.3~meV and $\sim$310~fs, respectively; the latter is estimated by performing TRPES on a highly oriented pyrolytic graphite (HOPG). The high repetition rate is suited for achieving high signal-to-noise ratio in TRPES spectra, thereby facilitating investigations of ultrafast electronic dynamics in the low pump fluence ($p$) region. TRPES of polycrystalline bismuth (Bi) at $p$ as low as 30~nJ/mm$^2$ is demonstrated. The laser source is compact and is docked to an existing TRPES apparatus based on a 250-kHz Ti:sapphire laser system. The 95-MHz system is less prone to space-charge broadening effects compared to the 250-kHz system, which we explicitly show in a systematic probe-power dependency of the Fermi cutoff of polycrystalline gold. We also describe that the TRPES response of an oriented Bi(111)/HOPG sample is useful for fine-tuning the spatial overlap of the pump and probe beams even when $p$ is as low as 30~nJ/mm$^2$. 
\end{abstract}

\pacs{}

\maketitle

\section{Introduction}
\label{Intro}

A non-equilibrated state of matter triggered by an ultrashort light pulse is acquiring great interest from both fundamental and application points of view~\cite{11PhysRep_Gamaly}. The impact by the pulse can induce a variety of phenomena including coherent oscillations, ultrafast phase transitions, and femtosecond laser ablations that are applicable to micromachining~\cite{08NPhoton_Mazur}, thin film growth~\cite{13RepProgPhys_Balling}, and clinical surgery~\cite{09JBioPhoton_Mazur}. The pulse may drive a solid state into warm-dense matter~\cite{04PRL_WarmDenseAu,09Science_WarmDenseAu,VailionsGamaly_NCom11,15NCom_Gd}, the extreme temperature and pressure of which mimic the conditions in the core of planets and stars~\cite{99Sci_Guillot}. The out-of-equilibrium phenomena can be studied by femtosecond pump-and-probe methods~\cite{Zewail}, in which a probe pulse snapshots the sample impacted by a pump pulse. 

Time-resolved photoemission spectroscopy (TRPES) has become a powerful tool to investigate the non-equilibrium properties of solid state matter from an electronic perspective~\cite{Haight_94,Haight_SurfSciRep95,PetekOgawa,Heinzmann_RSI01,Lisowski04,Carpene_RSI09,11OE_ARTEMIS_Frassetto,11RSI_Eberhardt,Perfetti_RSI12,Lanzara_TrPES_RSI12,12JJAP_Oguri,RSI13_HHG_Frietsch,14JES_Aeschlimann,Yb_RSI14,Ishida_RSI14,RSI15_HELIOS,Cilento_JES16}. In TRPES, a probing pulse has the photon energy that exceeds the work function, so that it can generate photoelectrons through a one-photon process: The distribution of the photoelectrons in energy and angle carries the information of the electronic structures. Nowadays, the investigations are mostly done at the pump fluence $p$ of $>$10~$\mu$J/cm$^2$ based on Ti:sapphire pulsed lasers~\cite{Nature03_Keller} operating at 10$^0$\,-\,10$^4$~kHz repetition. These include: ultrafast perturbation/melt of charge-density waves accompanying coherent oscillations~\cite{Perfetti_TaS2,Schmitt,TiS2e_Nature11,Ishizaka_TaS2,12NCom_TimeDomainClass,16NCom_CDW}; ultrafast modification~\cite{14PRL_UO2} and phase transition~\cite{VO2_Yoshida,14PRL_Wegkamp_VO2} of correlated insulators; investigations into graphene~\cite{Girez,Someya_APL}, graphitic materials~\cite{96PRL_graphite,Moos,Ishida_HOPG,14PRL_Bilayer_Hofmann,15PRB_Graphite_Bauer}, semiconductors~\cite{Azuma}, and cuprate superconductors~\cite{Perfetti_TTM,Bovensiepen,Lanzara_Science,Ishida_SciRep16}; disclosure of novel states~\cite{Gedik_Science13,Zhu_SciRep15}, dynamics~\cite{12PRL_Sobota,12PRL_WangGedik,Marsi_NCom14}, and functions~\cite{Ishida_SmB6,Neupane_PRL15} in topological insulators; and others~\cite{14PRL_UO2}. 
 
One of the natural pathways to understand the far-from-equilibrium phenomena is to approach them from the near-equilibrium, or mildly non-equilibrated states induced by a weak pump pulse. Besides, dynamics induced by a low pump fluence can be interesting on its own~\cite{13PRL_Higgs,13NMat_CDW_Fluct_Gedik}. In order to discern the small variations induced by the low-fluence pump in TRPES, it becomes practical to achieve higher repetition rate, or increase the number of pump-probe events per unit time, and improve the signal-to-noise ratio (S/N) of the dataset. Note, there is a limit in improving S/N by increasing the probe intensity. When too intense, the probe pulse generates a bunch of photoelectrons that repel each other through space-charge effects, thereby resulting in an undesired broadening in their distribution. 

Fiber lasers have emerged as a powerful high-repetition femtosecond light source~\cite{97APB_NelsonRev,08LasPhotRev_Wise,10JOptB_Richardson,13NPhoton_UltrafastFiberRev}. Optical fibers doped with rare-earth ions can be a lasing medium~\cite{61PRL_Snitzer,73APL_Nd_StoneBurrus,85_NdOscillation,86_ErOscillation,88_YbOscillation}, and oscillators made thereof can be mode locked~\cite{90_ElLett_1stFemtosecond_Nd,92IEEE_HoferNd,93OptLett_Tamura,97OptLett_CautaertsHanna,06OptExp_Andy} to generate pulses as short as sub-30~fs~\cite{08OptExp_28fs,12OptLett_21d6fs} at the typical repetition rate higher than $\sim$20~MHz~\cite{15PhotRes_19MHz}. Amplification for generating high-photon-energy pulses can also be done by doped fibers~\cite{97IEEE_YbAmplifier,11OptExp_Tunnermann,12OptLett_Constant,14NPhoton_Tunnermann} as well as by others such as external cavities~\cite{Gohle_Nature05,Pupeza_NaturePhoton13,Ozawa_10MHz}. The amplified pulses can be strong enough to be non-linearly converted up to 108~eV at 78~MHz~\cite{Pupeza_NaturePhoton13}. The fiber lasers thus foresee the high-repetition-rate ultrafast spectroscopy in the deep-to-extreme ultraviolet region~\cite{11OptExp_Tunnermann,12OptLett_Constant,14NPhoton_Tunnermann,Gohle_Nature05,12Nature_EUVComb,Pupeza_NaturePhoton13,13PRA_Ozawa,Ozawa_10MHz}; already, they have been nourishing the ultrafast spectroscopy in the low-photon-energy regions such as THz~\cite{11NPhoton_Kampfrath_NiO} and multi-photon-photoemission methods~\cite{14NPhys_CuiPetek}. Besides, fiber lasers are compact because of the flexibility, stable due to the all-solid nature, and cost effective owing to the economics scale of the telecommunications industry. 

The paper describes what we believe is the first TRPES apparatus based on a fiber laser system. One of the design concepts was to achieve a high repetition by utilizing Yb-doped fibers. We adopted a mode-locked Yb:fiber oscillator and a Yb:fiber amplifier for generating 1.2-eV pump and 5.9-eV probe pulses at the 95-MHz repetition. Non-linear crystals were used for generating the deep-ultraviolet probe, so that the laser system is all solid in nature. The repetition rate thus achieved is higher than the $\leq$80~MHz of the Ti:sapphire-based TRPES apparatus~\cite{12PRL_Sobota,Lanzara_TrPES_RSI12}. In addition, the energy resolution of 11.3~meV is comparable to 10.5~meV achieved in a 250-kHz TRPES apparatus~\cite{Ishida_RSI14,Ishida_SciRep16} which took considerations into the uncertainty principle; that is, to our knowledge, the Yb:fiber-based apparatus is the second to demonstrate the sub-20-meV resolution in TRPES. We show that the combination of the high repetition and high energy resolution is advantageous in detecting the subtle spectral changes induced by a weak pump fluence. Besides, the Yb:fiber laser system is compact and transportable, so that it can be easily installed into an existing TRPES apparatus, as we shall demonstrate herein. 

The paper is organized as follows. After the present introduction (Section \ref{Intro}), we describe the setup and performance of the apparatus in Section \ref{Setup}: The Yb:fiber laser system is sketched in \ref{layout}; The energy (11.3~meV) and time resolutions ($<$310~fs) are described in \ref{SpaceMirror} and \ref{TimeR}, respectively. The mitigation of the space-charge effects is also described in \ref{SpaceMirror}. In Section \ref{TrPES}, we present TRPES datasets of bismuth (Bi): TRPES at $p$ as small as 30~nJ/mm$^{2}$ is demonstrated in \ref{PolyBi} for a polycrystalline Bi film; In \ref{BiFilm}, we show the results of a highly-oriented Bi film grown on graphite, the pump-and-probe response of which turned out to be useful for cross-correlating the pump and probe beams in real time even at the low pump fluence. Finally, discussions are presented in Section~\ref{Conclusion}. In Appendix~\ref{appendix_SpaceMirror}, we present discussions of the mirror-charge effects that manifest when the fluence of the probe is low; In Appendix~\ref{appendix_Bi}, we describe the method to fabricate the highly-oriented Bi thin films.

\section{Setup and specifications}
\label{Setup}

\subsection{TrPES layout}
\label{layout}

\begin{figure}[htb]
\begin{center}
\includegraphics[width=7.8cm]{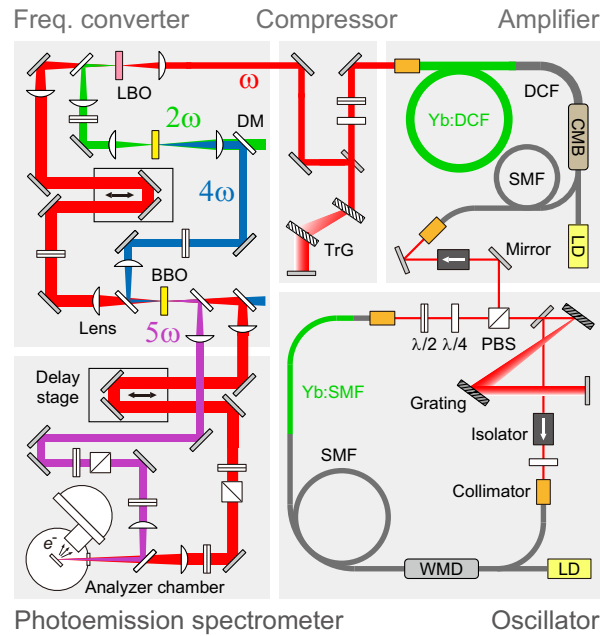}
\caption{\label{fig_setup} 
Schematic of the 95-MHz Yb:fiber laser based TRPES apparatus. LD: laser diode; WDM: wavelength division multiplexer; PBS: polarized beam splitter; DM: dichroic mirror; SMF: single-mode fiber; DCF: double-clad fiber; TrG: transmission gratings; $\lambda$/2 and  $\lambda$/4: half and quarter wave plates, respectively.}
\end{center}
\end{figure}

The layout of the TRPES setup is sketched in
Figure~\ref{fig_setup}. The laser system consists of an oscillator, an amplifier, a compressor, a frequency converter, and a section where the delay between the pump and probe pulses is controlled. 

The 95-MHz oscillator is a ring cavity composed of a Yb-doped single-mode fiber pumped by a 976-nm diode laser~\cite{08OptExp_28fs}. Mode locking occurs passively due to the non-linear polarization evolution. A pair of gratings optimizes the group-delay dispersion in the cavity. The oscillator delivers pulses with a central wavelength of 1062~nm, a bandwidth of 22~nm, and an average power of $\sim$20~mW. The seeding pulses are amplified in a Yb-doped double-clad fiber pumped by a 976-nm diode laser. Note, the fiber amplification is done without any reduction in the repetition rate, which is contrasted to the regenerative amplifications often adopted in Ti:sapphire laser systems. We amplify the seed up to $\sim$3~W, or up to $\sim$30~nJ/pulse, in order to generate sufficiently high flux of the fifth-harmonics probe; see later. The amplified output is then compressed by using a pair of transmission gratings. 

\begin{figure}[htb]
\begin{center}
\includegraphics[width=7.8cm]{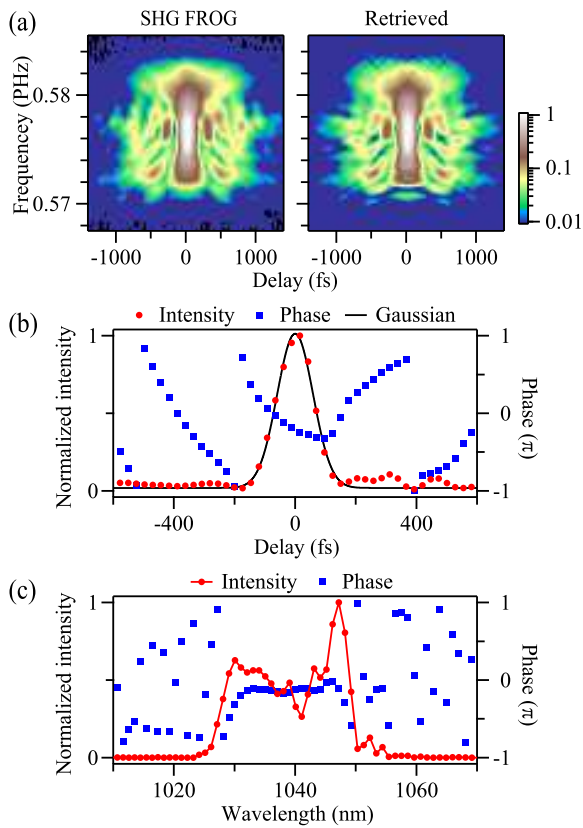}
\caption{\label{fig_FROG} 
Laser pulse characterized by the SHG-FROG method. (a) Experimental (left) and typical retrieved (right) SHG-FROG spectrograms. The FROG error~\cite{FROG_Terbino} of the retrieved image was 5.34\,$\times$\,10$^{-6}$. The retrieved intensity and phase of the pulse in time (b) and frequency (c) domains. }
\end{center}
\end{figure}

The amplified-and-compressed pulse was characterized by using the second harmonic generation (SHG) frequency-resolved optical gating method (FROG). The left and right panels in Figure~\ref{fig_FROG}(a) display the experimental and retrieved SHG-FROG spectrograms, respectively; the latter was generated by using a standard FROG algorithm~\cite{FROG_Terbino}. The retrieved intensity and phase of the pulse in time and frequency domains are shown in Figures~\ref{fig_FROG}(b) and \ref{fig_FROG}(c), respectively. The full width at half the maximum (FWHM) of the pulse is 140~fs, as estimated through fitting the profile by a Gaussian function [Figure~\ref{fig_FROG}(b)], and the central wavelength is 1040~nm [Figure~\ref{fig_FROG}(c)].  

The frequency up-conversion of the fundamental $\omega$ into the fifth harmonics 5$\omega$ is done by using two $\beta$-BaB$_2$O$_4$ (BBO) and a LiB$_3$O$_5$ (LBO) non-linear crystals. $\omega \to 2\omega$ is done by the LBO type I, and $2\omega \to 4\omega$ is done by the BBO type I. Finally, 4$\omega$ and $\omega$ (remainder after the first BBO) are mixed at the second BBO type I. The output of 5$\omega$ is as high as 0.3~mW. The thicknesses of the LBO, first BBO, and second BBO crystals are 2.0, 0.1, and 0.1~mm, respectively. The focal lengths of the lens before the LBO and first BBO are both 30~mm, while those focusing the 4$\omega$ and $\omega$ onto the second BBO are 75 and 100~mm, respectively. The beam after the second BBO is split into 5$\omega$ (probe) and $\omega$ (pump) by using a dichroic mirror, and the two beams are sent into the pump-and-probe delay section.  

The Yb:fiber laser system is docked to an ultra-high vacuum ($>$2 $\times$ 10$^{-11}$~Torr) chamber equipped with a hemispherical electron analyzer that has an acceptance angle of $\pm$18$^{\circ}$ along the entrance-slit direction (Scienta-Omicron, R4000). The analyzer chamber is shared by a TRPES apparatus based on a Ti:sapphire laser system delivering 1.5-eV pump and 5.8-eV probe pulses at 250-kHz repetition~\cite{Ishida_RSI14}. Because the probe beam of the Yb:fiber and Ti:sapphire systems happens to fall into the same photon energy region, the probe beam line is also shared by both systems. The arrival of the pump and probe pulses at the sample position is controlled by a delay stage inserted in the pump beam line. The pump and probe beams are finally joined at a dichroic mirror, and directed into the analyzer chamber through a CaF$_2$ window. The spot sizes of the pump and probe beams at the sample position can be measured and tuned by using a pin hole located inside the analyzer chamber~\cite{Ishida_RSI14}. 

The oscillator is enclosed in a box having a footprint of 20 $\times$ 20~cm$^{2}$. The amplifier, compressor, frequency converter, and part of the delay stage section are compacted on a board of 45 $\times$ 60~cm$^{2}$. The fiber laser system was developed and characterized on a vibration-free table and then transferred to a table next to the analyzer chamber; that is, the desk-top-sized laser system is transportable.

\subsection{Space-charge effects and energy resolution}
\label{SpaceMirror}

The space-charge broadening is an unwanted effect in photoemission spectroscopy, and has to be taken care of particularly when the probe light is pulsed. The flux of the probe beam has to be lowered until the spectral broadening becomes sufficiently small. Here, taking advantage that the apparatus is equipped with both 95-MHz Yb:fiber and 250-kHz Ti:sapphire laser systems, we demonstrate that, when using the same amount of the probing photons per second, the 95-MHz system is less prone to the space-charge effects compared to the 250-kHz system. We also describe the energy resolution of the apparatus. 

\begin{figure}[htb]
\begin{center}
\includegraphics[width=8.6cm]{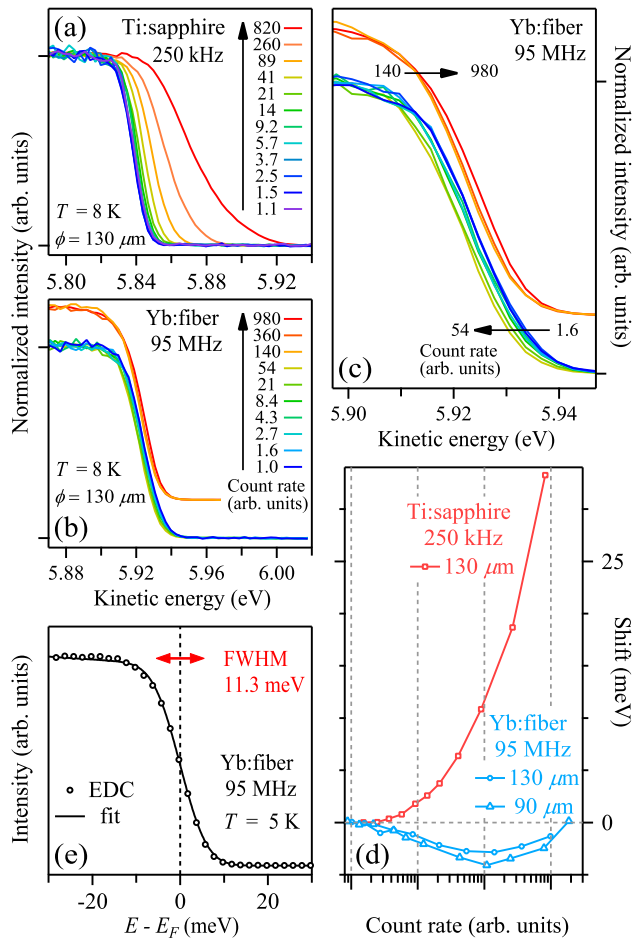}
\caption{\label{fig_SC} 
Space charge and mirror charge effects. (a, b) Fermi edge of gold recorded by the 250-kHz Ti:sapphire (a) and by the 95-MHz Yb:fiber (b) laser systems at various photoemission count rates. In (b), some of the spectra are vertically shifted for clarity. The diameter $\phi$ of the probe at the sample position was set to 130~$\mu$m. (c) Enlarged view of (b). (d) Shifts of the Fermi edge as functions of the photoemission count rate. The cases for $\phi$ = 90 and 130~$\mu$m are plotted for the 95-MHz setup. (e) Fermi edge of gold recorded by the 95-MHz Yb:fiber laser system. The fit (see text) shows the energy resolution of 11.3~meV. }
\end{center}
\end{figure}

When comparing the probe power dependency between the 250-kHz and 95-MHz systems, it becomes important to keep the conditions other than the repetition rate as constant as possible. First of all, the probing photon energy of the fiber-based system is higher only by 0.08~eV than that of the Ti:sapphire system. Therefore, the variations between the two setups in the matrix element, propagation velocity of the photoelectron bunch, and spectral window between the Fermi cutoff and work-function cutoff (Appendix~\ref{appendix_SpaceMirror}) can be regarded as marginal. In both setups, the spot diameter $\phi$ at the sample position was tuned to 130~$\mu$m by utilizing the pin hole attached next to the sample~\cite{Ishida_RSI14}. We also quickly switched the system from Ti:sapphire to Yb:fiber during the measurement in order to minimize the change of the condition of the sample surface as well as that of the analyzer. The whole dataset to be presented in Figure~\ref{fig_SC} was acquired within one day. Under these settings, we regarded the photoemission count rate from the gold sample as the measure of the probe-beam power, or the number of the probing photons per second, although in arbitrary units; see below. The data were acquired in a normal-emission geometry. The probe was $s$ polarized and the incidence angle was 45$^{\circ}$.

Figures~\ref{fig_SC}(a) and \ref{fig_SC}(b) respectively show the Fermi edge of polycrystalline gold recorded at various probe power values by the 250-kHz and 95-MHz systems. We took the photoemission count rate at 50 $\pm$ 15~meV below the Fermi level ($E_F$) as a measure of the probe power that was typically in the sub-mW range. The edge recorded by the 250-kHz system [\ref{fig_SC}(a)] broadened and shifted into higher kinetic energy upon increasing the power because of the space-charge effects. Apparently, such a shift-and-broadening is small in the dataset recorded by the 95-MHz system [\ref{fig_SC}(b) and \ref{fig_SC}(c)]. The results demonstrate that the space-charge effects are less pronounced in the high-repetition 95-MHz system. The probe-power-dependent shifts of the Fermi edge recorded by the 250-kHz and 95-MHz systems are summarized in Figure~\ref{fig_SC}(d): The shifts were estimated by fitting the edge by Fermi-Dirac functions convoluted with Gaussian functions. 

We here note that the low probing photon energy of $\sim$6~eV is advantageous for reducing the space-charge effects when the number of the photoelectrons consisting the bunch is concerned. At  $\sim$6~eV, only the electrons bound near the valence-band maximum can be emitted as photoelectrons, because the spectral cutoff due to the work function is located in the valence-band region; see Appendix~\ref{appendix_SpaceMirror}. Therefore, the number of the valence electrons that can be photo-emitted is intrinsically small at $\sim$6~eV compared to the cases for higher photon energies. Qualitatively, the suppression of the space charge effects at low probing photon energy is acknowledged in time-resolved multi-photon photoelectron spectroscopy~\cite{PetekOgawa}, although quantitative comparison to the present results is difficult because the photon energy of the pulses as well as other conditions is different in general. 

Looking closer into the dataset recorded by the 95-MHz system [\ref{fig_SC}(c)], we observe that, upon increasing the probe power, the edge initially shifts into low kinetic energies before shifting into high kinetic energies. When the spot diameter $\phi$ was reduced from 130 to 90~$\mu$m, the magnitude of the initial shift into lower kinetic energies was increased, as shown in Figure~\ref{fig_SC}(d). The shift into low kinetic energy upon increasing the probe power is most likely attributed to mirror-charge effects~\cite{Zhou} that manifested when the space-charge effects were reduced. For further discussion, see Appendix~\ref{appendix_SpaceMirror}.

We estimate the energy resolution of the apparatus by recording the Fermi edge of gold. Figure~\ref{fig_SC}(e) shows the spectrum recorded at $T$ = 5~K. The width of the entrance slit was set to 0.2~mm. Here, the probe power was reduced so that the space- and mirror-charge effects were small. By fitting the spectrum to a Fermi-Dirac function convoluted with a Gaussian function, the energy resolution [FWHM of the Gaussian] was estimated to be 11.3 $\pm$ 0.3~meV. The energy resolution is comparable to that of the 250-kHz Ti:sapphire-based system (10.5~meV)~\cite{Ishida_RSI14}, whose design concept was to achieve a high energy resolution by considering the uncertainty principle: A compromise has to be done in the time resolution when pursuing the energy resolution.

\subsection{Time resolution}
\label{TimeR}

We here describe the time resolution of the apparatus. The resolution can be estimated by performing TRPES on a highly oriented pyrolytic graphite (HOPG). We utilize the rise of the spectral intensity upon the arrival of the pump pulse, which occurs quasi-instantaneously when seen at the $\sim$100-fs resolution~\cite{Ishida_HOPG}. 

\begin{figure}[htb]
\begin{center}
\includegraphics[width=8.6cm]{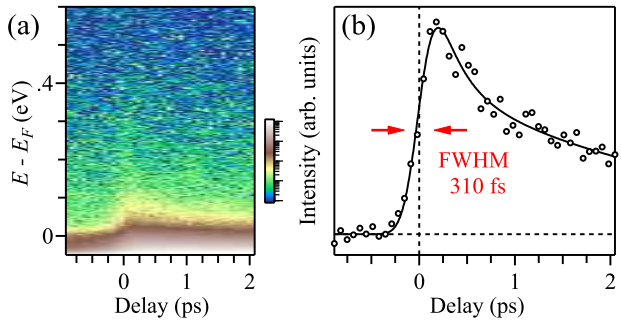}
\caption{\label{fig_HOPG} Time resolution estimated by TRPES of graphite. (a) Logarithmic spectral intensity mapped in energy and delay plane. (b) Spectral-weight variation at [0.02, 0.20~eV] as a function of delay. The fit (see text) shows the time resolution of 310~fs. }
\end{center}
\end{figure}

The HOPG sample was cleaved  by the Scotch-tape method (see Appendix~\ref{appendix_Bi}) in a vacuum chamber for sample preparation. Then, the sample was directly transferred to the analyzer chamber without exposing the sample to air. The TRPES dataset was recorded at room temperature. The delay stage of the pump was repetitively scanned until a sufficient S/N was achieved in the whole dataset~\cite{Ishida_RSI14}. Figure~\ref{fig_HOPG}(a) shows a logarithmic spectral intensity mapped in energy and delay plane. Upon the arrival of the pump pulse, the spectral intensity is spread into the unoccupied side, and then shows recovery. In Figure~\ref{fig_HOPG}(b), we plot the variation of the intensity in the energy window [0.02, 0.20~eV] as a function of the pump-probe delay $t$. By fitting the temporal profile to a two-exponential function convoluted with a Gaussian as done in Ref.~\cite{Ishida_HOPG}, the time resolution (FWHM of the Gaussian) is estimated to be 310~fs. 

\section{Demonstration of TRPES}
\label{TrPES}

Bi has played a pivotal role in the history of solid-state physics, and also continues to attract both scientific and industrial attentions~\cite{Fuseya_2014,Behnia_Science07}. Being an exemplary semi-metal, Bi is the first matter whose Fermi surface was revealed experimentally~\cite{Bi_FS39}. Studies of the ultrafast phenomena in Bi induced by femtosecond pulses shed light into coherent oscillations~\cite{Zeiger92,Hase02} and ultrafast bond-breaking dynamics~\cite{Lindemann,BondSoftening_Science07}. In addition, Bi single crystals exhibit novel metallic surface states, as revealed by angle-resolved photoemission spectroscopy (ARPES) studies~\cite{Ast_PRL01,Hofmann_Rev}. Since Bi is a heavy element, valence electrons are subjected to large spin-orbit interaction that results in intriguing phenomena: (1) Surface states of Bi exhibit Rashba-type spin splittings~\cite{Koroteev_PRL04}. (2) Bi adatoms can turn the surface of an ordinary semiconductor into exotic metal that shows giant Rashba splittings~\cite{GierzAst_BiSi,HattaBiGe_09,Yaji}. (3) A bi-layer Bi is a candidate of a two-dimensional topological insulator, owing to the the band inversion induced by the spin-orbit interaction~\cite{Murakami,12PRL_BiBilayer,14PRB_Yeom_BiBilayer}. (4) A three-dimensional topological insulator was realized in Bi alloyed by antimony~\cite{Hsieh_Nature}, a materials system also well known for its high thermoelectric performance. 

Here, we show TRPES datasets of Bi films recorded by the apparatus. First, we show the TRPES dataset of a polycrystalline Bi film at the low pump fluence of 30~nJ/mm$^2$, and demonstrate the high S/N achieved by the high repetition rate. Then, we show the time-resolved ARPES (TARPES) dataset recorded on the 111 face of Bi~\cite{Ast_PRL01,Hofmann_Rev,Koroteev_PRL04}: The surface electrons exhibited a giant pump-and-probe response, which turned out to be useful for tuning the spatio-temporal overlap of the pump and probe beams in real time even when the pump fluence is as low as 30~nJ/mm$^2$.

\subsection{TRPES of polycrystalline Bi}
\label{PolyBi}

A polycrystalline Bi film was prepared by evaporating Bi on a copper plate in vacuum. The TRPES dataset was recorded at room temperature. The diameter of the pump and probe beams were 250 and $<$150~$\mu$m, respectively, which were calibrated by using the pin-hole method~\cite{Ishida_RSI14}. The pump fluence $p$ was set to $\sim$30~nJ/mm$^2$. The dataset were acquired in a normal-emission geometry. The probe was $s$ polarized and the incidence angle was 45$^{\circ}$.

\begin{figure}[htb]
\begin{center}
\includegraphics[width=8.6cm]{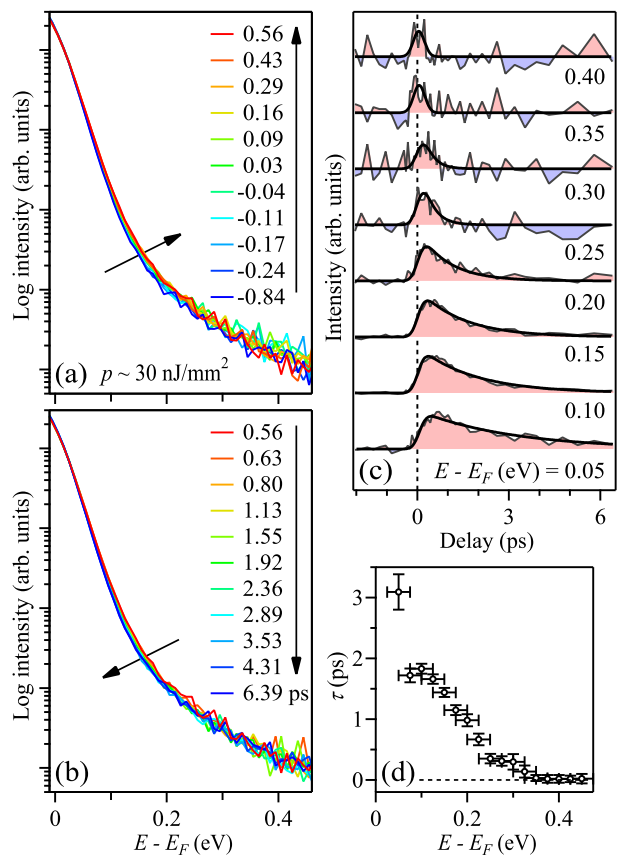}
\caption{\label{Fig_PolyBi} 
TRPES of polycrystalline Bi at the pump fluence as small as 30~nJ/mm$^2$. (a, b) Spectra recorded at $t \leq$ 0.56 (a) and $\geq$ 0.56~ps (b). (c) Pump-induced variation of the spectral intensity at various energies. The energy window was set to  $\pm$25~meV. Fitting functions (see text) are overlaid. (d) Decay time as a function of energy. The vertical bars indicate $\pm$1 standard deviation, and the horizontal bars indicate the energy window of $\pm$25~meV. }
\end{center}
\end{figure} 

Figures~\ref{Fig_PolyBi}(a) and \ref{Fig_PolyBi}(b) respectively show the TRPES datasets of $t \leq$ 0.56 and $\geq$ 0.56~ps. Upon the arrival of the pump pulse, the intensity in the unoccupied side increases [\ref{Fig_PolyBi}(a)] and then recovers gradually [\ref{Fig_PolyBi}(b)]. One can discern the pump-induced variation in the spectra as small as 10$^{-5}$ of the intensity at $E_F$, thanks to the high repetition rate of the Yb:fiber laser enabling the high S/N ratio. 
 
Figure~\ref{Fig_PolyBi}(c) shows pump-induced variation of the spectral intensity at various energies in the unoccupied side ($E - E_F > 0$). Clearly, the recovery slows on approach to $E_F$. The recovery time $\tau$ of the carriers is estimated by fitting the temporal profile at each energy by an exponential function $\exp(-t/\tau)$ convoluted by a Gaussian representing the time resolution. $\tau$ as a function of $E - E_F$ is plotted in Figure~\ref{Fig_PolyBi}(d). $\tau$ increases when $E$ approaches to $E_F$. We thus succeed in detecting the energy-dependent carrier dynamics induced by the pump fluence as small as 30~nJ/mm$^2$.

\subsection{TARPES of Bi(111) grown on graphite}
\label{BiFilm}
 
Here we present the TARPES dataset of Bi thin film grown on a HOPG substrate. Vacuum-evaporated Bi on the surface of HOPG forms Bi micro-crystals with the 111 face oriented normal to the surface~\cite{Ishida_RSI14,05SurfSci_Brown,10SurfSci_Brown}. This enabled us to investigate the pump-induced response of the Rashba-split surface states on Bi(111)~\cite{Koroteev_PRL04,Hofmann_Rev}. For the method to fabricate the thin film, see Appendix~\ref{appendix_Bi}.

\begin{figure}[htb]
\begin{center}
\includegraphics[width=8.6cm]{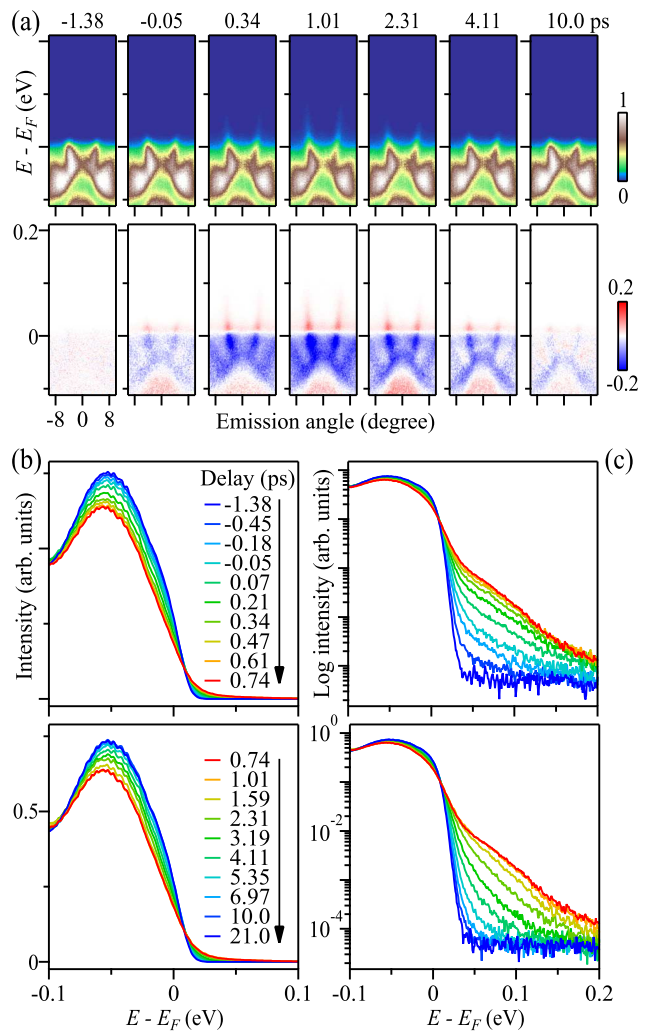}
\caption{\label{fig_BiTrAR} 
TARPES of highly-oriented Bi(111) grown on graphite. (a) TARPES images. The upper panels show band dispersions recorded at various delay-time values, and the lower panels show the images of the pump-induced difference. (b, c) Spectra averaged over the emission angle of $\pm$10$^{\circ}$. The upper and lower panels respectively show the averaged spectra recorded at $t$ $\leq$ 0.74 and $\geq$ 0.74~ps. Linear and logarithmic scales are adopted for the intensity axes in (b) and (c), respectively. }
\end{center}
\end{figure} 

TARPES of Bi(111)/HOPG was done at $T$ = 10~K at a pump fluence of $p\sim$ 30~nJ/mm$^2$. The upper and lower panels of Figure~\ref{fig_BiTrAR}(a) respectively display TARPES images and their difference to the averaged image before pumped. In the images, dispersions of the Rashba-split surface states are observed around the normal emission (emission angle 0$^{\circ}$), which is consistent to those presented in Ref.~\cite{Ishida_RSI14}. Upon the arrival of the pump pulse, the spectral intensity is spread into the unoccupied side, and then shows recovery. The spectra averaged over the emission-angle region of $\pm$10$^{\circ}$ are shown in Figures~\ref{fig_BiTrAR}(b) and \ref{fig_BiTrAR}(c); the former (latter) adopts linear (logarithmic) scale in the intensity axis. 

The pump-induced variation into the unoccupied side (Figure~\ref{fig_BiTrAR}) is surprisingly larger than that observed on the polycrystalline Bi film evaporated on a copper plate (Figure~\ref{Fig_PolyBi}). The signal appearing in the unoccupied side is easily observed in real time on the screen monitoring the multi-channel plate where the photoelectrons are detected. The large pump-and-probe response of the surface states of  Bi(111) is found to be very useful in maximizing the spatial overlap of the pump and probe beams at the sample position even when the pump fluence is as low as $\sim$30~nJ/mm$^{2}$ ($=$ $\sim$3.0~$\mu$J/cm$^{2}$), which corresponds to the pump power of 140~mW for the 250-$\mu$m beam diameter and 95-MHz repetition. If $p$ could be raised to $\sim$50~$\mu$J/cm$^{2}$, which is a level easily achieved by the 250-kHz Ti:sapphire-based system~\cite{Ishida_HOPG}, we could have used the response seen in the spectrum of HOPG as the real-time monitor for aligning the beams~\cite{Ishida_RSI14}. However, such a high pump fluence is not reachable by the present $\sim$3-W Yb:fiber laser system, because it corresponds to the pump power of $\gtrsim$2~W. Moreover, such a high load would heat up the sample. While HOPG is established as the reference sample for characterizing the pump and probe beams in the 250-kHz system~\cite{Ishida_RSI14}, we find Bi(111)/HOPG as a counterpart in the 95-MHz system. Besides, B(111)/HOPG is easy to fabricate and has high surface stability, as described in Appendix~\ref{appendix_Bi}: These characteristics facilitated us to adopt B(111)/HOPG as the reference sample for TRPES at low $p$.

\section{Discussion}
\label{Conclusion}

We described the TRPES apparatus based on an all-solid Yb:fiber laser system delivering 1.2-eV pump and 5.9-eV probe at the 95-MHz repetition. The energy and time resolutions were 11.3~meV and 310~fs, respectively. The high repetition as well as the high energy resolution facilitated us to discern the subtle spectral changes induced by a weak pump pulse, as demonstrated by the TRPES of polycrystalline Bi film: The spectral intensity variation of 10$^{-5}$ to the intensity at $E_F$ was detected.  

Concerning the investigations into the non-equilibrated state of solid state matter, the high repetition in the pump-and-probe method works most nicely when the pump fluence is low. If a high fluence pump was done at a high repetition, the sample would suffer from heatings and ablations. The general perspective is that the high (low) repetition is suited for investigating the weakly (intensively) pumped states. Thus, the high-repetition TRPES achieved by the implementation of Yb:fibers complements the Ti:sapphire-based TRPES operating at lower repetition rates. The use of the high-repetition-rate Yb:fiber laser system resulted in the improved S/N in the TRPES datasets and expands the pathways to perform TRPES in the very low excitation limit, the conditions of which are of major importance for the studies close to equilibrium that might reveal dynamics otherwise hidden or masked at stronger excitation conditions. The setup may be suitable for investigating novel dynamics mildly induced by a weak pump, such as the oscillations in superconductors in its superconducting phase~\cite{13PRL_Higgs} and those in cuprates in its pseudogap phase~\cite{13NMat_CDW_Fluct_Gedik}. 

We also demonstrated the compatibility of the Yb:fiber laser system to the Ti:sapphire-based TRPES. First of all, the all-solid Yb:fiber laser system is compact and transportable. This enabled us to set the system next to the Ti:sapphire laser system. Besides, a variety of optics components can be shared by the two systems, because the photon energies of the probes are similar. Moreover, the methods as well as concepts developed for operating Ti:sapphire-based TRPES measurements~\cite{Ishida_RSI14} were readily applicable. These include: the pin-hole method for directing the pump and probe beams to the sample position at a controlled beam size; the method to visualize the infrared pump and ultraviolet probe beams simultaneously; the know-hows to stabilize the laser, temperature, and sample position; the repetitive delay scanning during the data acquisition; and to use not-too-short pulses to keep a fair resolution in energy. A method developed uniquely for Yb:fiber TRPES was to use Bi(111)/HOPG as a reference sample. The giant response of the surface states on Bi(111) enabled us to cross-correlate the pump and probe pulses at the sample position in real time even when the pump fluence was as low as $p \sim$ 30~nJ/mm$^2$. 

We hope that the descriptions would facilitate Yb:fiber TRPES to be an accessible option to those based on Ti:sapphire lasers, and would expand the pathways for investigating the ultrafast phenomena in matter.

\begin{acknowledgments}
This work was supported by Photon and Quantum Basic Research Coordinated Development Program from MEXT, NEDO, and JSPS KAKENHI No.~26800165 and No.~15K17675. 
\end{acknowledgments}

\appendix

\section{Space-charge and mirror-charge effects}
\label{appendix_SpaceMirror}

\begin{figure}[htb]
\begin{center}
\includegraphics[width=7cm]{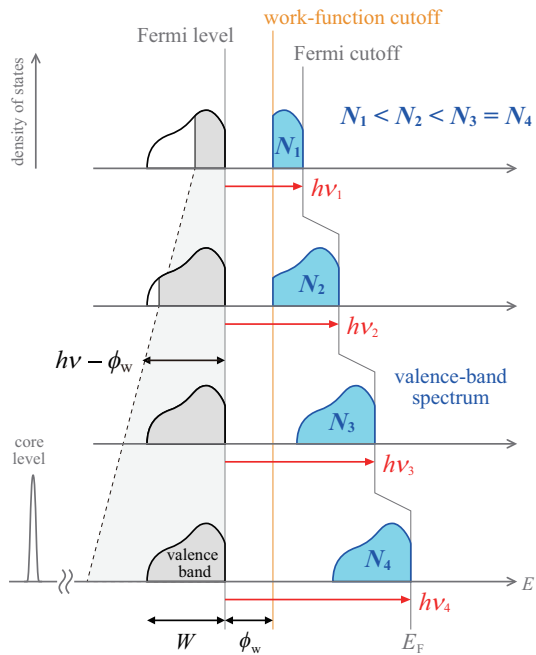}
\caption{\label{fig_PhotonEng} 
Density of states,  photon energy of the probe $h\nu$, and spectral cutoff due to the work function. 
The work-function cutoff occurs in the valence-band region when $h\nu$ is low and fulfills the condition $h\nu - \phi_w < W$ (upper two panels). 
Here, $\phi_w$ ($\sim$5~eV) is the work function, and $W$ ($\sim$10~eV) is the width of the valence band in energy. 
 As $h\nu$ is increased (from top to bottom), the gray shaded energy window $h\nu - \phi_w$ expands, 
 and the Fermi cutoff shifts away from the work-function cutoff in the valence-band spectrum. 
 $N_i$ is the number of the valence electrons that can contribute to form the photoelectron bunch when the probing photon energy is $h\nu_i$. 
 $N_i$ increases until $h\nu_i > \phi_w + W \sim$ 15~eV is reached (lower two panels; note, $N_1 < N_2 < N_3 = N_4$). }
\end{center}
\end{figure}

The shift and broadening of the spectrum as a function of the probing photon flux occur due to the combination of the space- and mirror-charge effects~\cite{Zhou}. The effects depend on a variety of parameters such as~\cite{Zhou,ChulkovEchenique}: (1) the number of the photoelectrons $N$ forming a bunch; (2) the pulse length; (3) the size and shape of the probe beam; (4) the energy distribution of the photoelectrons that depends on the valence band structure of the sample and photon energy of the probe; and (5) the conductivity of the sample. 
Concerning the shift, the space-charge and mirror-charge effects, respectively, cause an upward and downward shift of the Fermi cutoff in kinetic energy; the latter occurs because the photoelectron bunch is effectively attracted by its own mirror charge induced in the metal sample. The dominance of the downward shift of the cutoff in the low-flux region, as seen in Figures~\ref{fig_SC}(c) and \ref{fig_SC}(d), can be understood by considering $N$ $\to$ 1: In the limiting case, there is no space-charge effects but only the mirror-charge effects. To date, however, only upward shifts have been reported when the duration of the probing pulses are in the femtoseconds~\cite{Bauer_SpaceCh,Lanzara_SpaceCh}; downward shift is reported only when the pulse duration is $\sim$60~ps~\cite{Zhou}. Thus, the observation of the downward shift resolves the mystery why the signatures of mirror-charge effects were not observed for the TRPES probe in the femtoseconds. The successful observation presumably owes to the followings: (1) The photon energy of the probe is as low as 5.9~eV, so that only the electrons bound near $E_F$ can contribute to the photoelectrons; that is, $N$ is intrinsically small compared to the cases when the probes have higher photon energies; see Fig.~\ref{fig_PhotonEng}. (2) The high energy resolution of $<$20~meV and high S/N facilitated us to discern the downward shift as small as 1~meV. The observation of the downward shift in femtosecond TRPES nicely connects into the regime of mutli-photon photoelectron spectroscopy where the mirror-charge effects dominate over the space-charge effects~\cite{HoferFauster,EcheniquePendry} because it usually deals with very low-kinetic-energy photoelectrons generated by the low-fluence pulses. 

\section{Fabrication of oriented Bi thin film on HOPG}
\label{appendix_Bi}

\begin{figure*}[htb]
\begin{center}
\includegraphics[width=14cm]{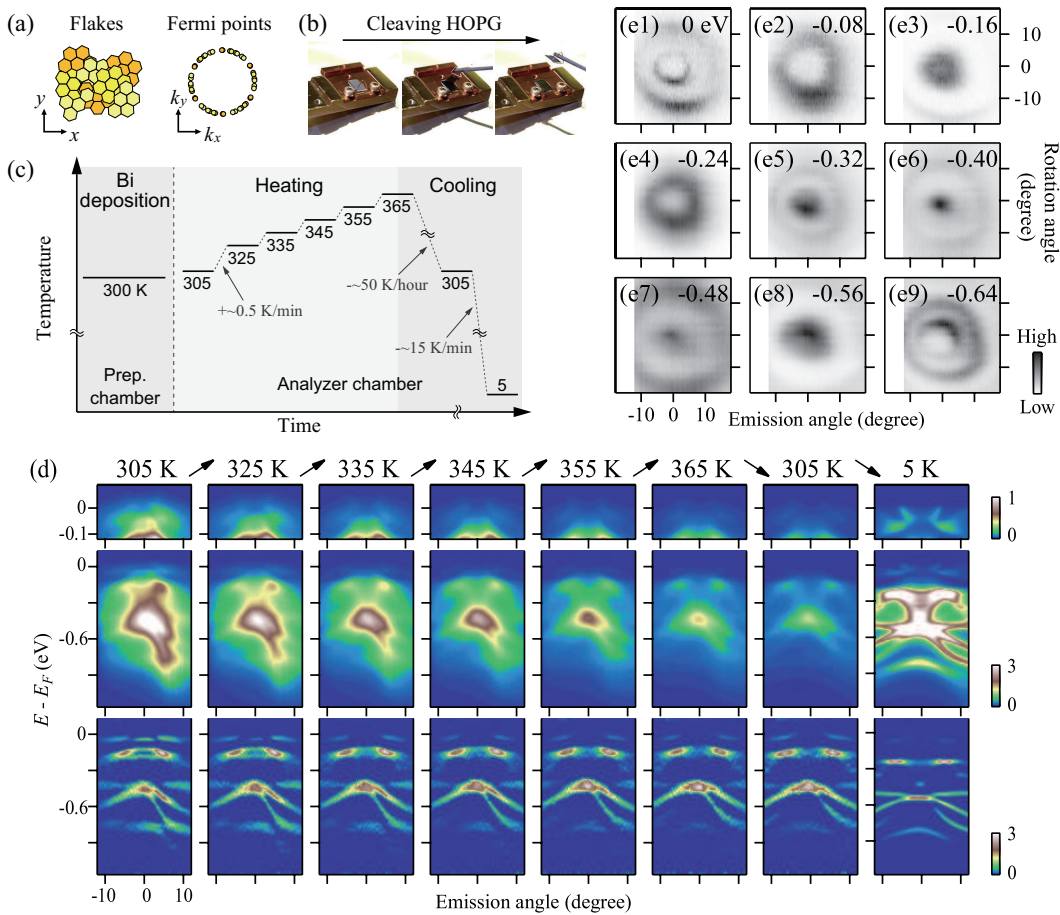}
\caption{\label{fig_BiGrowth} 
Growth of highly-oriented Bi(111) on HOPG. (a) The schematic of the HOPG structure. Graphite micro-crystals are randomly oriented in the $x$-$y$ plane. The corresponding Fermi points of HOPG form a Debye-ring-like distribution in momentum ($k_x$-$k_y$) space. (b) Cleaving of HOPG by the Scotch-tape method. (c) Evaporation of Bi and subsequent heating and cooling. (d) ARPES images of Bi film monitored during the heating-and-cooling procedure. The upper and middle panels respectively show the ARPES images near $E_F$ and in a wide energy range. The lower panels are the second derivative of the ARPES images with respect to energy. (e) Angular distributions of photoelectrons from the Bi film grown on HOPG at various energies. }
\end{center}
\end{figure*}

We describe the method to obtain the highly-oriented Bi(111) micro-crystalline thin film. Vacuum-evaporated Bi on cleaved surface of HOPG forms micro-crystals with the 111 face oriented normal to the surface~\cite{Ishida_RSI14,05SurfSci_Brown,10SurfSci_Brown}. Because HOPG consists of micron-sized graphite sheets randomly oriented in plane as shown in Figure~\ref{fig_BiGrowth}(a), the Bi micro-crystals grown on HOPG also have random in-plane orientation, as we describe below. 

First, we obtain a clean surface of HOPG by using the Scotch-tape method~\cite{Novoselov_Science}. The tape attached on HOPG is peeled off in the vacuum chamber for sample preparation. The vacuum can be at the level of $\sim$5 $\times$ 10$^{-8}$~Torr. In Figure~\ref{fig_BiGrowth}(b), we show the snapshots taken during the removal of the tape. Note, the demonstration is done in air. To our experience, the tape is compatible to ultrahigh vacuum unless baked. The Scotch-tape method nicely works as long as the cleaving is done around the room temperature. At cryogenic temperatures, the tape hardens, so that it becomes less successful to have a good cleave. 

Next, we evaporate Bi on the exposed surface of HOPG. The evaporator consists of an alumina crucible heated by a tungsten filament, and is equipped in the preparation chamber. The typical thickness of the film, after $\sim$30-min evaporation, is more than 1000~\AA, as judged by the color of the copper sample holder deposited by Bi. The vacuum level can be of the order of $\sim$5 $\times$ 10$^{-8}$~Torr during the deposition; that is, an ultra-high vacuum of $\lesssim$10~$^{-9}$~Torr is not necessary.

Subsequently, the sample is transferred into the analyzer chamber without exposure to air. Here, we anneal the sample following a procedure described in Figure~\ref{fig_BiGrowth}(c). The sample is heated by using a ceramic heater, whose main usage is to control the temperature of the sample during ARPES measurements. In Figure~\ref{fig_BiGrowth}(d), we show ARPES images monitored during the heating-and-cooling sequence. The image recorded on the as-deposited sample [the left-most column in Figure~\ref{fig_BiGrowth}(d)] already shows bands typical to Bi(111) surface, although some other features also exist; see later. Upon raising the temperature across $\sim$340~K, the bands in the image become sharp, indicating that the film is annealed. Similar annealing occurs around 340~K in a Bi film deposited on Si(111)-7\,$\times$\,7 surface~\cite{Bian_ARPES}. We heat the sample up to 365~K to ensure a homogeneous annealing. Then the sample is cooled down to 305~K. One can see that some features in the image are removed after the annealing, such as the band occurring near $E_F$ around the normal emission (emission angle around 0$^\circ$). As the sample is further cooled down to 5~K, the bands in the image become even more sharp [right-most column of Figure~\ref{fig_BiGrowth}(d)].

In Figure~\ref{fig_BiGrowth}(e), we show constant-energy maps of the angular distributions of the photoelectrons emitted from the Bi film grown on HOPG. A Debye-ring-like distribution is observed in the maps. This shows that the film consists of Bi micro-crystals whose 111 faces are oriented normal to the surface, while their orientations being random in plane. The highly oriented structure is similar to that of HOPG, as shown in Figure~\ref{fig_BiGrowth}(a). The hexagonally-warped band dispersions of Bi(111) are thus averaged out in the azimuth to result in the ring-shaped photoelectron distributions in the maps. 

We find that the surface of the Bi film thus obtained is extraordinary stable. The bands in the ARPES image were observable even after the sample had been kept in the preparation chamber for two months. Even when the surface had acquired residual gas in the vacuum chamber at cryogenic temperatures, the sharp bands in the ARPES image were recovered after raising the sample temperature to 300~K. However, once the sample was exposed to air, the bands were not  observable any more. 

The high stability of the surface, besides the ease in the sample preparation, make the highly-oriented Bi(111)/HOPG a convenient reference sample for TRPES. We use the TARPES signal of the film when searching for and fine-tuning the spatio-temporal overlap of the pump and probe pulses at the focal point of the electron analyzer, as described in Sec.~\ref{BiFilm}.


%

\end{document}